\documentstyle[12pt,aps,colordvi]{revtex}
     \newcommand{\be}{\begin{equation}}
     \newcommand{\ee}{\end{equation}}
     \newcommand{\bea}{\begin{eqnarray}}
     \newcommand{\eea}{\end{eqnarray}}
     
\begin{document}
     \draft
     \title{\bf{Fluctuation Induced Non-Fermi Liquid Behavior near
     a Quantum Phase Transition in Itinerant Electron Systems. }} 
     \author{Suresh G. Mishra and P. A. Sreeram} 
     \address{Institute of Physics, Bhubaneswar 751005, India} 
     \maketitle 
     \begin{abstract}

 The signature
     for a non-Fermi liquid behavior near a quantum phase 
     transition has been observed in thermal and transport properties
     of many metallic systems at low temperatures. In the present work we
     consider specific examples of itinerant ferromagnet as well as
     antiferromagnet in the limit of vanishing transition
     temperature. The temperature variation of spin susceptibility,
     electrical resistivity, specific heat, and NMR relaxation rates
     at low temperatures is calculated in the limit of infinite
     exchange enhancement within the frame work of a self consistent
     spin fluctuation theory. The resulting non-Fermi liquid behavior
     is due to the presence of the low lying critically damped spin
     fluctuations in these systems. The theory presented here gives
     the leading low temperature behavior, as it turns out that the
     fluctuation correlation term is always smaller than the mean
     fluctuation field term in three as well as in two space
     dimensions. A comparison with illustrative experimental
     results of these properties in some typical systems has been
     done. Finally we make some remarks on the effect of disorder
     in these systems.

     \end{abstract} 
     \vspace{.1in}

     \section{Introduction}    
     The description of electronic contribution to the low
     temperature behavior of metals in terms of the Fermi liquids has
     been highly successful. \cite{Noz64} The low lying excitations of
     the Fermi liquid manifest themselves in various thermodynamic
     and transport properties, such as the specific heat varying as
     $C_v = \gamma T $, a temperature independent (Pauli) spin
     susceptibility, $ \chi = 2 \mu_B N(\epsilon_F) $, where
     $N(\epsilon_F)$ is the density of states at the Fermi energy, a
     temperature dependent electrical resistivity varying as $
     \Delta \rho \sim A T^2 $, and a linearly temperature dependent
     NMR relaxation rate. $ T_1^{-1} \sim T $ (Korringa). The
     values of the these coefficients, such as $\gamma$, $A$, etc.,
     however, are material dependent. For some transition metals these
     are about one order of magnitude larger than in normal metals,
     and in some compounds containing a large concentration of
     rare-earth or actinide elements such as Ce, Yb or U, these
     values are about a thousand times larger, particularly the value
     of $\gamma$ and the zero temperature susceptibility. \cite{Lee86}  

     The normal Fermi liquid behavior as mentioned above is
     understood within the Landau phenomenological theory, where the 
     effect of interaction in a Fermi system is expressed in terms of
     a few parameters which renormalize the physical quantities with
     respect to their free Fermi gas values. For example, the
     modifications in specific heat, spin susceptibility and
     isothermal compressibility are given by, 
     $ {C_v}/{C_v^0} = \frac {m^*}{m} = 1 + (F_1^s/3) $,  
     $ {\chi}/{\chi^0} = (m^* /m)/(1 + F_0^a) $, and $ {\kappa_s}/
     {\kappa_s^0} = (m^* /m)/ (1 + F_0^s) $ respectively. (The
     superscript $ 0 $ denotes the free Fermi gas values, other
     notations are standard). \cite{Noz64}  
     The basic reason for the success of the Landau theory is the
     largeness of quasi-particle life time near the Fermi surface,
     i.e., $ \tau^{-1} \sim \mid \epsilon \mid ^2 \ll \epsilon $,
     where $ \epsilon = (E-E_F)/E_F $. From these relations it is
     clear that for a certain value of the Landau parameters (i.e.
     $F_0 $, and $ F_1$), the corresponding quantities become very
     large, which in turn may indicate a neighborhood of certain
     phase transition. For example, $ F_0^a \rightarrow -1 $ implies 
     magnetic instability, or $ F_0^s \rightarrow \infty $, a
     condensation. In the present work we consider the Fermi system
     in the vicinity of such a transition and seek an explanation of
     the non-Fermi liquid behavior of certain substances in this
     regime. \cite{JPCM96} It seems the Fermi liquid theory gives
     indication of the incoming electronic phase transition as the
     coupling constant is changed, but it does not consider the
     effect of incipient fluctuations in a self-consistent manner.  

     There are many examples of electronic phase transitions where
     the coupling constant tunes the transition. These are known as
     the quantum phase transition. For example $ 1 - UN(\epsilon_F) >
     0 $ gives instability towards ferromagnetism, $ 1 - U \chi(Q) >
     0 $, gives antiferromagnetic instability corresponding to a wave
     vector $ Q $, and $n^{1/3}a_H > 0.26$ describes the metal
     insulator transition due to Coulomb correlation as suggested by
     Mott. These are essentially zero temperature transitions,
     however, in general, $T_c << T_F $, where $T_F$ is the Fermi
     temperature of the system. In contrast the classical phase
     transition occurs at finite temperature and is described by
     balance in the energy needed (loss) to create disorder with gain
     in entropy due to disorder such that the Free Energy, $ F = U
     -TS $, is reduced. One more difference is that the statics and
     dynamics become correlated in quantum phase transition.
     \cite{He76,Mill93} This is principally due to non-commutativity
     of various terms in the Hamiltonian. For example, consider the
     Hubbard model, 
     \begin{equation}
     H = \sum_k \epsilon_k n_k + U \sum_i n_{i\uparrow}n_{i\downarrow},
     \label{hubbardm}
     \end{equation}
     for correlated electrons. Here the kinetic energy and the $U$
     terms do not commute. (Otherwise the model will be trivial to
     solve). Technically, this means that one should introduce
     ``time'' and the Feynmann time ordering in the functional
     integral for the partition function. The order parameter field
     becomes ``time dependent''. The time variable thus acts as an
     extra dimension. This leads to a change in the critical
     behavior. \cite{He76} At a first glance it seems that the 
     critical behavior would be the same as that of a $D+1$
     dimensional classical system. However, detailed analysis shows
     that the critical behavior (or the Upper Critical Dimension)
     depends on the dispersion and damping of the order parameter
     fluctuations. The reason is that the spin susceptibility for a
     ferromagnet above $T_c$ is given by, \cite{Izu63} 
     \begin{equation}
     \chi(q,\omega^+) \approx \frac{N(\epsilon_F)} 
     {(1-UN(\epsilon_F)) +\delta q^2 - \frac{i\gamma\omega}{q}}.
     \label{ferrochi}
     \end{equation}
     For free electron gas $\gamma =1/12$ and $\delta = 1/2$.
     At $T_c$, $ 1-UN(\epsilon_F) \rightarrow 0 $, and therefore
     $\omega \approx q^3$, gives the order parameter dispersion. In
     case of antiferromagnetism, the staggered spin susceptibility is
     given by, \cite{Ued77} 
     \begin{equation}
     \chi(Q+q,\omega^+) \approx \frac {\chi^0(Q)} {1-U\chi^0(Q) + 
     \delta q^2-i\gamma\omega}
     \label{afchi}
     \end{equation}
     In this case, $\omega \approx q^2$ at the critical point. A
     dynamical exponent $z$ is introduced, which reflects the change
     in the static critical behavior. In particular the scaling
     dimension of the quartic interaction is given by $\epsilon =
     4-(d+z)$ with $z=3$ for ferromagnets and $z=2$ for
     antiferromagnets. \cite{He76} In field theory $z=1$, since
     $\omega$ and $q$ are linearly related, and have the same scaling
     form. At present the application of the renormalization group to
     quantum critical phenomenon in particular the correlation of the
     static and dynamic behavior is a subject of intense activity. We
     refer the reader to \cite{He76,Mill93,Sach95,Son97} for detailed
     discussion. To summarize, the vicinity to the phase transition
     point and the fermionic nature of correlated electronic system
     undergoing a phase transition change the nature of the phase
     transition itself as well as the Fermi liquid behavior expected
     in this system. The reason for this behavior is the smallness of
     the transition temperature, $T_c$ compared with the Fermi
     temperature $T_F$. This aspect gets revealed more clearly as
     $T_c \rightarrow 0$.  
          
     To calculate various physical properties, we take specific
     examples of ferromagnetic and antiferromagnetic transition in
     itinerant electron system in two as well as three dimensions in
     the limit of vanishing transition temperature near the
     transition temperature. These two examples represent two
     different types of quantum critical behavior. The basic reason
     is that in ferromagnet the order parameter is a conserved
     quantity, while in the antiferromagnet it is not. This
     difference is reflected in the dispersion of their respective
     order parameter fluctuation as shown in Eqns. (\ref{ferrochi})
     and (\ref{afchi}). The microscopic calculation is done within
     the self consistent spin fluctuation theory developed earlier by
     Ramakrishnan and one of us \cite{TVR74,MR78a,MR78b,MR85} among
     many others \cite{MD72,MK73,LT85}. For details of the spin
     fluctuation theory we refer the reader to the monograph by
     Moriya \cite{Mo85}. A brief review is given in \cite{SGM90}. We
     first briefly review the spin fluctuation theory and then write 
     expressions for spin susceptibility, resistivity, specific heat
     and the nuclear magnetic relaxation rate. Similar expressions
     for staggered susceptibility and other quantities in
     antiferromagnets are also written. These quantities are then 
     calculated in the limit of large exchange enhancement (i.e. in
     the limit of $ \chi_{P} \chi(T = 0)^{-1} \equiv \alpha (0) 
     \rightarrow 0 $). Though $T_c = 0$, fluctuation effects are
     observable well above $T_c$. The temperature dependence need not
     be Fermi-Liquid-like because of the low lying fluctuation
     (Bosonic) degrees of freedom. 

     \section{Spin Fluctuation Theory} The basic motivation for
     constructing the spin fluctuation theory is the largeness of the 
     susceptibility (Stoner) enhancement factor $ 1/\alpha (0) $.
     In such case a highly paramagnetic system at low temperature can
     be considered to be in the vicinity of a magnetic transition.
     The temperature variation of various physical quantities is
     therefore governed by transverse and longitudinal spin
     fluctuations. Though the order parameter vanishes above the
     transition, the effect of fluctuations is observable well above
     the transition. There are many equivalent formulations available
     on this idea \cite{Mo85,SGM90}. A recent one is due to
     McMullan,\cite{M87} which is based on functional integration
     using Grassmann variables and collective coordinate 
     transformation. We briefly summarize our approach and then
     compile results on some physical properties.  

     Consider the Landau expansion for the free energy $F(M,T)$,
     in power of the order parameter $M $, viz. 
     \be
     F(M,T)=F(0,T)+{1 \over 2} A(T)M^2+{1 \over 4} BM^4-HM 
     \label{landaufe}
     \ee
     where $H$ is the field conjugate to $M$. The temperature
     dependence of various quantities in this theory arises due to 
     $A(T)$ and $B$. For example, the spin susceptibility for the
     paramagnetic phase is given by,
     \be
     \chi^{-1}(T)=A(T).
     \label{landauchi}
     \ee
     ($A(T)$ and $\alpha (T) $ have qualitatively the same
     temperature dependence and differ only by some numerical
     factors, e.g.  $ A(T) = \alpha (T) / 2 N(\epsilon_F) $ for 
     ferromagnets, which we ignore and identify $ A(T) $ with 
     $\alpha (T) $ now onwards.) Similarly the magnetization in the
     ordered phase is, 
     \be
     M^{2}(T)=-{\alpha(T)\over{B}},
     \label{landaumsq}
     \ee
     and the equation of state is given by, 
     \be
     {H\over{M}} =\alpha(T)+BM^{2}.
     \label{landauhbym}
     \ee
     The expansion coefficients $\alpha(T)$ and $B$ have been calculated
     in various approximation schemes. In the Ginzburg Landau theory
     for classical phase transition, $\alpha(T)$ is taken as $(T - T_c)$
     and $B$ as independent of temperature. This leads to the Curie
     Weiss law for the susceptibility and the well known mean field
     critical exponents. In the mean field theory of itinerant
     ferromagnet, 
     \be
     \alpha_{MF}(T)= \biggl(1- U N(\epsilon_F)\biggr),
     \label{ferroalmf}
     \ee
     and $B$ is again a constant. In this case the temperature
     dependence of physical quantities near $T_c$ comes from that of
     integral over density of states through a Sommerfeld expansion.
     It is weak, of the order of $T^{2}/T_{F}^{2}$, and therefore it
     does not give a Curie-Weiss form for the spin susceptibility.
     This issue is tackled in the spin fluctuation theory, where
     $\alpha(T)$ is given by,\cite{MR78a,MR78b} 
     \be
     \alpha_{SF}(T)= \alpha (0) + u_{4} (2D^T+3D^L).
     \label{ferroalsf}
     \ee
     Here $ \alpha (0) $ is the susceptibility enhancement factor at
     $T=0$. This includes the mean field part $\alpha_{MF}(T) $ and
     the zero temperature part of the fluctuation self energy whose
     finite temperature part comprises the second term. Here $D^T$
     and $D^L$ are transverse and longitudinal spin fluctuation
     amplitudes obtained by the internal frequency summation in the
     diagrams shown in Fig.\ (\ref{feyndiag1}) (a,b,c). The main
     contribution to the temperature variation of various physical
     quantities is governed by these amplitudes. The factor $u_4$ in
     the second term is is a dimensionless short range four
     fluctuation coupling constant obtained after integration over
     fast fermionic degree of freedom.
     
     The above result has been derived microscopically, within the
     functional integral scheme on a model of interacting electrons.
     We consider Hubbard model as applied to itinerant
     ferromagnets and for brevity consider only spin degrees of
     freedom. Applying the Stratanovich- Hubbard functional integral
     transformation the partition function can be written as 
     \bea 
     Z &=& \mbox {tr} \int \prod_{q,m} \frac{d\xi _{q,m}}{\pi } \exp [- \sum
     _{q,m} \mid \xi_{q,m} \mid^2  \nonumber \\
      & & -\int _{0}^{\beta } du [\sum _k \epsilon _k n_{k,\sigma
     ,u}-(\frac{U}{\beta })^{1/2}\sum _{q,m} (\xi _{q,m}^*.S_{q,m} \exp
     (z_m u) + h.c.)]] 
     \label{hspartfunc1}
     \eea
     where $\xi _{q,m}$ is the spin fluctuation field of wave vector
     $q$ and frequency $z_m$ ($=2\pi im/ \beta $). Also, $ \epsilon_k
     $ is the kinetic energy of the electrons and $U$ denotes a short
     range inter atomic repulsion. Integrating over
     the electronic degrees of freedom, we have free energy
     functional $F(\xi _{q,m})$ for interacting spin fluctuations;
     that is 
     \be
     Z=\int \prod_{q,m} \frac{d\xi _{q,m}}{\pi } 
     \exp [{-\beta F(\xi _{q,m})}] . 
     \label{hspartfunc2}
     \ee
     Parameters of this model, e.g. fluctuation spectrum, fluctuation
     coupling vertices are determined by properties of the underlying
     fermion system. Since these parameters (e.g. the Stoner
     enhancement factor for ferromagnets or the staggered
     susceptibility for antiferromagnets) are such that spin
     fluctuations are low lying excitations, this transformation is
     specially helpful for an analysis of temperature dependent
     properties of weak itinerant electron ferromagnets and
     antiferromagnets. The free energy functional $F({\xi _{q.m}})$~
     then expanded in powers of these fluctuation fields up to a
     quartic term and a self consistent mean fluctuation field
     approximation (quasi harmonic approximation, or the self
     consistent renormalization scheme of Moriya) can be generated.
     The mean fluctuation field approximation corresponds to the
     diagram (a), (b) and (c) in Fig.\ (\ref{feyndiag1}), and shown
     in a compact manner in Fig. (\ref{feyndiag2}) where the double
     wiggle represents the dressed propagator $D({\bf q})$. The
     details are given in earlier papers \cite{MR78a,MR78b}.  One can
     also estimate corrections due to higher order fluctuation terms.
     Figs.\ (\ref{feyndiag1}) (d) and (e) and (f) represent typical
     higher order fluctuation correlated terms.  
     
\section{Physical Properties Near Quantum Critical Point}
     \subsection{Spin Susceptibility} 
     The self consistent equation for the temperature dependence of
     $\alpha(T)$ is given by Eq.\ (\ref{ferroalsf}) which is written 
     explicitly as  
     \be
     \alpha(T) = \alpha(0) + \lambda \sum_q \int d\omega n(\omega ) 
     {\rm Im}\chi(q,\omega^+)
     \label{ferroalstart}
     \ee
     where, $\lambda $ is related to $U_4$, $ n(\omega ) =
     {(e^{\omega/T}-1)}^{-1} $, is the Bose distribution function,
     and 
     \be
     \chi(q,\omega)=\frac{N(\epsilon_F)}{\alpha (T) + \delta q^2 - \imath
     \frac{\pi\omega\gamma}{2q}}, 
     \label{ferrochiqomega}
     \ee
     is spin susceptibility for the ferromagnetic case. 
     ($\omega$ and $T$ are written in units of $\epsilon_F$ and $ q $
     in units of $k_F$. We have taken $\hbar = 1$ and $k_B = 1$).
     Performing the frequency integral,  
     \be
     \alpha(T) = \alpha(0) + \frac{\lambda}{\pi}\sum_q 
     q \{\ln(y) - \frac{1}{2y} -\psi(y)\} 
     \label{ferroalpha1}
     \ee
     where,
     \be
     y = \frac{q}{\pi^2 \gamma T}(\alpha(T) + \delta q^2).
     \label{ydef}
     \ee
     An interpolation formula for,
     \bea
     \phi(y) & \equiv & \{\ln(y) - \frac{1}{2y} -\psi(y)\} \nonumber
     \\ & \simeq & \frac{1}{2y+12y^2}, 
     \label{phiyapprox}
     \eea
     which is valid for small as well as large $y$ is useful in
     calculating the momentum integral. For three dimension, 
     \be
     \alpha(T) = \alpha(0) + \frac{\lambda}{2\pi^3}\int \frac{q^3
     dq}{(2y+12y^2)}. 
     \label{ferroalpha2}
     \ee

     A finite $\alpha(0)$ introduces two regions of
     temperatures.\cite{MR78a}. For $T < \alpha(0)$ one gets the
     standard paramagnon theory results; and for $ \alpha(0) < T < 1
     $ one gets the classical Curie Weiss susceptibility,      
     \be
     \chi = \chi_P / \alpha(T) \simeq \mu_B^2/(a \alpha(0) + T).
     \label{ferroalphaltTchi}.
     \ee
     That is like susceptibility of a collection of classical spins.
     This feature gets revealed more clearly if we put $\alpha(0) = 0
     $ in the expression for $\chi (T) $ and solve the equation self
     consistently. In this case the paramagnon regime ($ T \le
     \alpha(0) $) shrinks to zero and and a classical behavior is
     expected down to $T=0 $. One is then essentially calculating
     susceptibility of a ferromagnet with $T_c= 0$. Since $\alpha(0)$
     is taken to be zero and there is only one region of temperature 
     $T < 1$. In this case, typical $y \le 1$, the limiting form is
     obtained using the form $ \phi(y) \approx \frac{1}{2y}$ ( valid
     for $y << 1$). We then find that, 
     \be
     \alpha (T) = \frac{T}{\delta}\Biggl[ q_T
     -\biggl(\frac{\alpha(T)}{\delta} \biggr)^{1/2} \arctan \biggl(q_T 
     \frac{\delta}{\alpha(T)}\biggr)^{1/2}\Biggr]
     \label{ferro3dalpha}
     \ee
     where, $q_T$ is a thermal cutoff such that $y_{q_T} \approx 1$.
     For the form of $y$ given by Eq.\ (\ref{ydef}), the estimate of the 
     cutoff is
     $q_T^3 \approx T\gamma/\delta$ or $q_T \approx T^{1/3}$. The
     dominating contribution to $\alpha(T)$ comes from the first term
     which is given by, $T^{4/3}$. However, since $\delta$ is small,
     the thermal cutoff $q_T$ is high $\approx q_c$ (the spin
     fluctuation energy rises only slowly with $q$). Thus, $\alpha(T)$
     rises nearly linearly with $T$. This is the classical spin
     fluctuation behavior, first pointed out for itinerant
     ferromagnets by Murata and Doniach.\cite{MD72} Note that we have
     assumed $T < 1$, i.e. the system is degenerate. Even so, since the
     characteristic fluctuation energy $\alpha(0)$ is zero, the
     system behaves classically with regard to spin fluctuations. 
     An estimate of the size of the second term is obtained by
     putting $\alpha(T) \approx T^{4/3}$. We then find it to be of the
     order $T^{1/3}$ relative to the first term. Since $T^{1/3}$ is
     not very small, it is essential to do a self consistent
     calculation, particularly in the classical regime which is of
     interest in the present calculation.

     We have calculated $\alpha(T) $ and other properties in two
     dimension also. For this we consider the same approximate form
     of the spin susceptibility or the fluctuation propagator as in
     three dimension, the effect of dimensionality is considered 
     only through the phase space in the momentum integration. The
     assumption regarding the form of susceptibility function in two
     dimension is in doubt. It is well known \cite{Kit68} that the
     Lindhardt function from where this functional form has been
     derived has a different analytic form in two dimension. As far
     as the low momentum behavior is concerned the assumption is
     closer to the reality if $\delta $ is considered to be far
     smaller than its value in three dimension. For the sake of
     comparison we assume the same value of $\delta$ in three as well
     as in two dimension.  

     Following the same procedure as in three dimensions we find in
     two dimensions a logarithmic temperature dependence, 
     \begin{equation}
     \alpha(T)=\frac{T}{2\delta}{ \ln}\biggl(\frac{\delta {
     q_c^2}} {\alpha(T)}\biggr)
     \label{ferro2dalpha}
     \end{equation}

     Because of the Bose factor $1/(\exp (\omega /T) -1) $, the
     number of thermal (classical) fluctuations becomes smaller and
     smaller as $ T \rightarrow 0 $ (i.e. as the $T_c$ approaches).
     This reduces the phase space for the fluctuation correlations.
     In the RNG analysis of Hertz \cite{He76} and others \cite{Mill93} this 
     requires the introduction of a suitably scaled `energy' variable as a 
     degree of freedom additional to the three momentum variables. In effect
     the dimensionality increases, and the behavior becomes
     mean-field. We see this explicitly in our procedure of calculating
     the fluctuation correlation correction perturbatively. The terms
     involving two or more internal thermal spin fluctuations are
     shown in Fig.\ (\ref{feyndiag2}). These have been calculated in detail
     earlier.\cite{MR78a} It turns out that apart from a numerical
     factor the two internal thermal spin fluctuation term has the
     same temperature dependence as the mean fluctuation field term.
     However, the three internal thermal spin fluctuation term, $
     \approx T^2 \ln \biggl((1/3\alpha(T)\biggr)$ in 3D ferromagnet.
     We see that this term is of the order of $T{ \ln}T$ relative to
     the simplest non-vanishing contribution. The perturbation
     expansion therefore converges.
     
     For a finite $T_c$ ferromagnet the mean fluctuation field theory
     is valid outside the critical regime. As the critical regime
     approaches higher order fluctuation correlations become
     comparable to the mean fluctuation term. In the present case,
     in contrast the mean fluctuation field term gives the leading 
     critical behavior. The reason is the following. Suppose for
     $\alpha(0) = 0$, $\alpha(T) \sim T ^ \lambda $. Then the quantum
     region $T \ll \alpha(T) $ means $T^{1-\lambda} \ll 1 $ and it
     occurs only if $ \lambda \le 1 $. This is not possible, and so
     one always has the other classical (Curie - Weiss) 
     region. Here the fluctuation correlation term is of the form $
     T ^2 \ln (1/\alpha(T)) \sim T^2 \ln (1/T)^\lambda \ll 
     T^\lambda $. If $ \lambda \sim 1$, the correlation term never
     becomes more important than the mean fluctuation field term.
 
     In case of Antiferromagnets, the formalism is identical. One
     replaces the Pauli susceptibility with the staggered
     susceptibility for non interacting electron system $ \chi_0({\bf
     Q}) $, for brevity we retain the same notation for the
     enhancement factor which is defined in the present case as
     $\alpha (0) = \chi_0({\bf Q})/\chi({\bf Q}) $. The expansion of
     the dynamic staggered susceptibility, $\chi_0({\bf Q}+{\bf
     q},\omega)$ for small {\bf q} and small $\omega$ around the static 
     staggered susceptibility is also written in the form,
     \cite{Ued77} 
     \begin{equation}
     \chi ({\bf Q}+{\bf q},\omega^+) = \frac{\chi^0 ({\bf
     Q})}{\alpha(T)+\delta q^2 -\imath \gamma\omega}. 
     \label{afchiQomega}
     \end{equation}
     Making similar transformations as for the ferromagnetic case, we
     get, 
     \be
     \alpha(T) = \alpha(0) + \frac{\lambda}{2}\sum_q \frac{1}{(2z+12z^2)}
     \label{afalpha1}
     \ee
     where
     \be
     z = \frac{(\alpha(T) + \delta q^2)}{2\pi\gamma T}
     \label{zdef}
     \ee
     Thus, for 3D antiferromagnets,
     \be
     \alpha(T) = \alpha(0) + \frac{\lambda}{2\pi}\int 
     \frac{q^2 dq}{(2z+12z^2)}.
     \label{afalpha2}
     \ee
     The result turns out to be identical to the ferromagnetic case
     once we consider only the $z < 1 $, where the corresponding momentum
     cut off turns out to be $T^{1/2} $,  
     \begin{equation}
     \alpha(T)=\frac{T}{\delta}\Biggl[{
     q_c}-\sqrt{\frac{\alpha(T)}{\delta}} 
     \arctan \biggl(\sqrt{\frac{\delta}{\alpha(T)}}{ q_c}\biggr)\Biggr],
     \label{af3dalpha}
     \end{equation}
     and in two dimension there is again a logarithmic behavior,
     \begin{equation}
     \alpha(T)=\frac{T}{2\delta}{   \ln}\biggl(\frac{\delta {   q_c^2}}
     {\alpha(T)}\biggr).
     \label{af2dalpha}
     \end{equation}

     \subsection{Resistivity} The electrical resistivity for pure
     transition and rare earth metals is usually calculated within a 
     two band model, \cite{ML66} where the `conducting' electrons come from 
     s-band while the d-electrons contribute to magnetism. The d-band is
     assumed to be narrow and the d-electrons heavy. The conducting
     s-electrons scatter from the spin fluctuations corresponding to
     d- electrons. The temperature dependent part of the
     resistivity due to this mechanism for a 3d-ferromagnet is
     given by, \cite{Zim60,Math68,BeMo83}  
     \be
     \rho(T) \propto \frac{1}{T}\int q^3 dq \int
     Im\chi(q,\omega^+) \omega
     n(\omega) (1+n(\omega )) d\omega 
     \label{rhostart}
     \ee
     The frequency integral can be performed by first using the
     identity, 
     \be
     \frac{1}{T^2} \omega n(\omega) (1+ n(\omega )) =
     \frac{\partial}{\partial T}n(\omega) 
     \label{identity}
     \ee
     leading to,
     \begin{eqnarray}
     \rho (T) 
     && \approx T \int dq q^3 \int d\omega \frac{dn(\omega)}{dT} 
     \frac{q \omega}{q^2(\alpha(T)+\delta q^2)^2 + \omega^2} \nonumber \\
     && \approx \int dq~q^4 y \phi^{\prime} (y)  
     \label{ferrorhoapprox}
     \end{eqnarray}     
     where, $\phi^{\prime}(y)=\frac{d\phi(y)}{dy}$, and $\phi(y)$ and $y$
     are given by Eqns.\ (\ref{phiyapprox}) and (\ref{ydef}) respectively.
     In the limit $y <1 $ the momentum integral gives,  
     \begin{equation}
     \rho(T)=\frac{T}{2\delta}\Biggl[{
     q_c^2}-\frac{\alpha(T)}{\delta} { \ln}\biggl(\frac{\alpha(T) + \delta
     { q_c^2}}{\alpha(T)}\biggr)\Biggr].
     \label{ferro3drho}
     \end{equation}
     With $q_c \sim T^{1/3} $ we recover the well known result
     $\Delta \rho \sim T^{5/3} $.\cite{Math68} However, the self
     consistency correction changes the power of temperature.
     Similarly for two dimensions, 
     \begin{equation}
     \rho(T)=\frac{T}{\delta}\Biggl[{
     q_c}-\sqrt{\frac{\alpha(T)}{\delta}} 
     \arctan \biggl(\sqrt{\frac{\delta}{\alpha(T)}}{ q_c}\biggr)\Biggr].
     \label{ferro2drho}
     \end{equation}
     
     The case of 3D antiferromagnets formalism is similar except for
     the power of $ q $ in the momentum integral. This is due to the
     fact that the small momentum expansion is not done around $ q =
     0 $ but around $q = Q $, the antiferromagnetic wave vector. The
     result is, \cite{Ued77}
     \be
     \rho(T) \propto \frac{1}{T}\int q^2 dq \int Im\chi({\bf
     Q+q},\omega^+) \omega n(\omega) (1+ n(\omega )) d\omega .
     \label{afrho1}
     \ee
     Following the same steps as for the ferromagnetic case, we get, 
     \be
     \rho(T) \propto T \int q^2 dq \frac{(1+12z)}{2z(1+6z)^2}
     \label{afrho2}
     \ee
     where z is given by Eq.\ (\ref{zdef}) The result in the limit of
     $ z <1 $ 
     is, 
     \begin{equation}
     \rho(T)=\frac{T}{\delta}\Biggl[{
     q_c}-\sqrt{\frac{\alpha(T)}{\delta}} 
     \arctan \biggl(\sqrt{\frac{\delta}{\alpha(T)}}{ q_c}\biggr)\Biggr].
     \label{af3drho}
     \end{equation}
     Similarly for two dimensions, $ \rho(T)=\frac{T}{2\delta}{
     \ln}\biggl (\frac{\delta {q_c^2}} {\alpha (T)}\biggr). $
     
     \subsection{Specific heat} 
     The spin fluctuation contribution to the free energy within the
     mean fluctuation field approximation (or quasi harmonic
     approximation) is given by,\cite{MR85}
     \begin{equation}
     \Delta\Omega =\frac{3 T}{2}\sum_{q,m}\ln \{ 1-U\chi_{qm}^0
     +\lambda T \sum_{q^\prime,m^\prime}D_{q^\prime m^\prime} \}.
     \label{freeenegy1}
     \end{equation}
     Where $D_{q,m}$ is the fluctuation propagator which is related
     to inverse dynamical susceptibility, and $\chi_{qm}^0$ is the
     free Fermi gas (Lindhardt) response function.  The argument of
     the logarithm is related to inverse dynamic susceptibility.
     Considering only the thermal part of the integral and ignoring
     the zero point part, we perform the frequency summation and
     obtain, 
     \begin{equation}
     \Delta\Omega_{Thermal}=\frac{3}{\pi}\sum_q \int_0^\infty 
     \frac{d\omega}{e^{\omega/T}-1} \arctan \{\frac{\pi\omega/4q}
     {\alpha(T)+\delta q^2}\},
     \label{freeenegy2}
     \end{equation}
     Integrating over frequency, we get,
     \begin{equation}
     \Delta\Omega_{Thermal}=3 T \sum_q 
     \biggl( \ln\Gamma(y)-(y-\frac{1}{2})\ln(y)
     +y -\frac{1}{2}\ln (2\pi)\biggr).
     \label{freeenegy3}
     \end{equation}
     where, $y$ is given by Eq.\ (\ref{ydef}). Once the free energy
     correction is known, the specific heat correction is given by
     \begin{eqnarray} 
     \frac{\Delta C_v}{k_B} & = &
     -T\frac{\partial^2\Delta\Omega}{\partial T^2} 
     \nonumber \\
     & = & - 3T^2\sum_q \biggl[(\frac{2}{T}\frac{\partial y}{\partial
     T}
     +\frac{\partial^2 y}{\partial T^2})\phi(y)+(\frac{\partial
     y}{\partial T})^2
     \frac{\partial\phi(y)}{\partial y} \biggr] \nonumber \\
     & = & 6 \int q^2 dq \{\phi^\prime(y)(\frac{q}{\pi^2\gamma}
     \frac{\partial\alpha(T)}{\partial T}-y)^2 + T \phi(y)
     \frac{q}{\pi^2\gamma} \frac{\partial^2\alpha(T)}{\partial T^2}\}.
     \label{ferro3dspe1}
     \eea
     Making the small $y$ approximation and introducing the
     appropriate cutoff,
     \bea
     \frac{\Delta C_v}{k_B} & \approx &
     \frac{1}{2\pi^2}\biggl[(T^2\frac{\partial^2\alpha(T)}{\partial T^2}
     +2T\frac{\partial\alpha(T)}{\partial T} ) \int_0^{q_c} dq
     \frac{q^2}{\alpha(T)+\delta q^2} \nonumber \\
     ~& & -T^2\biggl(\frac{\partial\alpha(T)}{\partial T}\biggr)^2 
     \int_0^{q_c} dq \frac{q^2}{(\alpha(T)+\delta q^2)^2}
     -\int_0^{q_c} dq~q^2 \biggr] \nonumber \\
     & = & -\frac{1}{\delta}\biggl(T^2\frac{\partial^2\alpha(T)}{\partial T^2}
     +2T\frac{\partial\alpha(T)}{\partial T}\biggr)\Biggl[{   q_c} 
     -\sqrt{\frac{\alpha(T)}{\delta}} 
     \arctan \biggl(\sqrt{\frac{\delta}{\alpha(T)}} {
     q_c}\biggr)\Biggr] \nonumber \\ 
     &&
     +\frac{T^2}{2\delta}\biggl(\frac{\partial\alpha(T)}{\partial T}\biggr)^2
     \Biggl[\frac{1}{\sqrt{\alpha(T)\delta}} \arctan
     \biggl(\sqrt{\frac{\delta}{\alpha(T)}}{   q_c}\biggr)
     -\frac{{   q_c}}{\alpha(T)+\delta {   q_c^2}}\Biggr]
     +\frac{{   q_c^3}}{3}.
     \label{ferro3dspe2}
     \end{eqnarray}
     The last result is obtained after the momentum integration. 
     Approximately the terms can be arranged as,
     \be
     C_V \approx  \frac{1}{2\pi^2}
     \Biggl[ q_c^3 + T^2\biggl(\frac{\partial\alpha(T)}{\partial T}\biggr)^2 
     \frac{1}{\sqrt{\alpha(T)}} -
     \biggl(T^2\frac{\partial^2\alpha(T)}{\partial T^2} 
     +2T\frac{\partial\alpha(T)}{\partial T}\biggr)q_c \Biggr].
     \label{ferro3dapprox}
     \ee
     The first term gives the classical result (for constant cutoff),
     the second dominant term gives leading temperature correction,
     the last term is about two order of magnitude small at the
     temperature range of interest. Similarly in two dimensions,
     \bea
     C_V= &&
     -\frac{1}{2\delta}\biggl(T^2\frac{\partial^2\alpha(T)}{\partial T^2}
     +2T\frac{\partial\alpha(T)}{\partial T}\biggr){
     \ln}\biggl(\frac{\delta { q_c^2}}{\alpha(T)}\biggr) \nonumber \\ 
     && +
     \frac{T^2}{2\delta\alpha(T)}\biggl(\frac{\partial\alpha(T)}{\partial T}
     \biggr)^2\biggl(1+\frac{\alpha(T)}{\delta {   q_c^2}}\biggr)^{-1} 
     + \frac{{   q_c^2}}{2}
     \label{ferro2dapprox}
     \eea
     The calculation for antiferromagnet is identical except that $y$
     is replaced by $z$ in Eq.\ (\ref{ferro3dapprox}) and the final equation 
     in terms of $\alpha(T) $ turns out to be identical except that the
     temperature dependence of $\alpha(T) $ is different in an
     antiferromagnet.

     \subsection{Nuclear Spin Relaxation Rate}
     The nuclear spin relaxation rate in metals is given by the
     Korringa relation, \cite{Kor50}
     \begin{equation}
     \frac{1}{T_1T} \approx \biggl(\frac{\Delta H}{H}\biggr)^2
     \label{korringarel}
     \end{equation}
     which essentially tells that $1/T_1$ is proportional to the
     square of the static spin susceptibility of metals, which in
     turn, is independent of temperature for most normal metals. 
     However, it has been pointed out by Moriya \cite{Mor63} long ago
     that this relation gets modified in presence of electron
     correlations. The nuclear spin-lattice relaxation rate in
     metals is given by, 
     \begin{equation}
     \frac{1}{T_1T} \sim \sum_q \frac{{\rm Im} \chi^
     {-+}(q,\omega_0^+)}{\omega_0} 
     \label{relaxrate}
     \end{equation}
     where $\omega_0$ is the nuclear magnetic resonance frequency
     which is taken to be very small ($\rightarrow 0 $) in the problem of 
     nuclear spin relaxation rate. Substituting the expression for $
     \chi^ {-+}(q,\omega_0^+) $ and taking the limit, we have,
      \be
     \frac{1}{T_1T}  \sim \sum_q \frac{1}{q(\alpha(T)+\delta q^2)^2},
     \label{ferrorelax}
     \ee
     for a ferromagnet in three dimension. After the momentum
     integration, the result is $(T_1T)^{-1} \sim \alpha (T) ^{-1} $. For
     normal Fermi liquid $\alpha (T) $ is constant but in the present
     case it varies as $T^{4/3} $. This leads to a non-Fermi liquid
     behavior again. Similar calculation is done for antiferromagnets. 
   
     \subsection{Effect of Disorder} The effect disorder can be
     included in the above mentioned formalism by modifying the
     propagators and vertices in diagrams for the spin fluctuation
     self energy. This has been done in our earlier
     papers.\cite{SGMP92,SGMP93} In presence of disorder the electron
     moves in random way getting scattered from impurities
     repeatedly. This introduces a finite mean free path for
     electron, and a finite life time $\tau$ in the electron
     propagator, which also modifies the free particle-hole
     propagator (diffuson), free particle-particle propagator
     (Cooperon) and electron-spin fluctuation vertex. The correction
     to $ \alpha (T) $ to leading order is given by, 
     \be
     \alpha(T)=\alpha _{SF}(T)- \alpha _{d}(T),
     \label{disoalpha1}
     \ee
     where $\alpha _{d}(T)$ is correction due to diffusive modes. It
     is given to the leading order in $1/\epsilon_{F}\tau$ as 
     \be
     \alpha_{d}(T) \sim [1-\sqrt{(2\pi \tau T)} ]/ {(\epsilon_{F}\tau)^2 } , 
     \label{disoalpha2}
     \ee
     for $T\ll 1/\tau$ and it vanishes otherwise. Clearly the
     disorder introduces a new energy scale $(1/\tau )$, in the
     lowest temperature range as shown schematically in Fig. \
     (\ref{line}). In the case of non-vanishing $\alpha(0)$, in the
     case of ferromagnet in three dimension, the susceptibility
     inverse $ \alpha (T) $ behaves as $ \sim T^{1/2} $ for $ T<
     1/\tau $, as $ T^2/\alpha(0) $ for $ 1/\tau < T < \alpha (0) $,
     and as $ T^{4/3} $ for $ \alpha (0) < T < 1$.  Similarly the
     resistivity correction $ \Delta \rho (T) $ behaves as $ \sim
     T^{1/2} $ for $ T< 1/\tau $, as $ T^2/ \sqrt{\alpha(0)} $ for $
     1/\tau < T < \alpha (0) $, and as $ T^{5/3} $ for $ \alpha (0) <
     T < 1$. In the case of a zero-$T_c $ system the quantum
     fluctuation regime ($ T < \alpha(0) $) vanishes and the other
     two regimes merge. At lowest temperature the effect of
     diffusive mode seems to give the dominant contribution, i.e.
     $\alpha(T) \sim T^{1/2} $ and $\Delta \rho (T) \sim T^{1/2} $
     but a more detailed analysis is needed.  
   
     \subsection{Summary} The following table summarizes our
     results. In the first column Fermi liquid theory results are
     written, other columns compile the fluctuation theory results.
     These results are presented in three rows for each property. The
     first row gives results from a non-self-consistent calculation, i.e.,
     for example, when only the first term in Eq. (\ref{ferro3dalpha}) for 
     $\alpha(T)$ is considered but with a proper momentum cut off. This
     behavior is expected in the extreme low temperature range. These
     results are in general known, but presented in the coherent form
     for the first time. The second row gives these results with the
     temperature dependence of $\alpha(T) $ taken in account and the
     integration performed with the functional form for $\phi(y)$
     valid for all $y$ but approximated by Eq.\ (\ref{phiyapprox}). The 
     power of temperatures so obtained depends slightly on the
     temperature regime considered (i.e.  whether $T $ is in $10^{-3}
     - 10^{-2} $ or otherwise). The third row gives the classical spin
     fluctuation results, where the Bose factor $n(\omega)$ is
     approximated as $T/\omega$, (effectively the first row with a
     constant cutoff).  The experimental results are expected to lie
     between those given in rows one and three.

     \begin{tabular}{|c|c|c|c|c|c|}
\hline
~~ & ~~ & ~~ & ~~ & ~~ & ~~ \\
~~ & ~Fermi Liquid ~& ~Ferro (3D)~ & ~Antiferro (3D)~ &~Ferro (2D)~ &~ Antiferro (2D)~ \\
~~ & ~~ & ~~ & ~~ & ~~ & ~~ \\
\hline
~~ & ~~ & ~~ & ~~ & ~~ & ~~ \\
~~ & ~~  & $T^{4/3}$ & $T^{3/2}$ & $T{\rm ln}T$ & $T{\rm ln}T$ \\ 
$(\chi (T))^{-1} $ & Const. &  $T^{1.20}$ & $T^{1.44}$ & $T^{0.87}$ & $T$ \\ 
~~ &~~ & $T$  &   $T$  &  $T$   & $T$  \\ 
~~ & ~~ & ~~ & ~~ & ~~ & ~~ \\
     \hline
~~ & ~~ & ~~ & ~~ & ~~ & ~~ \\
~~ & ~~ & $T^{5/3}$ & $T^{3/2}$ & $T^{4/3}$  & $T{\rm ln}T$ \\ 
$\rho (T) $ & $T^2$ & $T^{1.56}$ & $T^{1.45}$ & $T^{1.24}$  & $T$ \\ 
~~ & ~~ & $T$  & $T$  & $T$  & $T$  \\ 
~~ & ~~ & ~~ & ~~ & ~~ & ~~ \\
     \hline
~~ & ~~ & ~~ & ~~ & ~~ & ~~ \\
~~ & ~~ & $T$ & $T^{3/2}$ & $T^{2/3}$  & $T$ \\ 
 $C_v (T) $ & $T$ &  $T^{0.74}$ & $T^{0.99}$ & $T^{0.52}$  & $T^{0.86}$ \\ 
~~ & ~~ & Const.  &  Const.  & Const. & Const  \\ 
~~ & ~~ & ~~ & ~~ & ~~ & ~~ \\
\hline
~~ & ~~ & ~~ & ~~ & ~~ & ~~ \\
~~ & ~~ & $T^{-4/3}$ & $T ^{-3/4}$ & 
$(T{\rm ln}T)^{-3/2}$ & $(T{\rm ln}T)^{-1}$ \\ 
$ (T_1T)^{-1} $ & Const. &  $T^{-1.284}$ & $T^{-0.72}$ & 
$T^{-1.305}$ & $T^{-1}$ \\ 
~~ & ~~ & $T^{-1}$ & $T^{-1/2}$  &  $T^{-3/2}$ & $T^{-1}$ \\ 
~~ & ~~ & ~~ & ~~ & ~~ & ~~ \\
\hline
\end{tabular}
\newpage
     \section{Experimental results:} In this section we give examples
     of materials exhibiting non-Fermi liquid behavior at low
     temperatures and also compare some results with a theory
     presented above. The most popular example of system showing
     non-Fermi liquid behavior is, of course, the high temperature
     superconductors \cite{MP94}. It seems, however, the effective low
     dimensionality, the specific nature of the density of states and
     structural aspect of the Fermi surface (nesting etc.) play
     important role in this system. We therefore want to consider
     examples from three dimensional correlated electronic system in
     the neighborhood of electronic phase transition.
     
     The next example is that of phosphorus-doped silicon (Si:P).
     \cite{PAA91} 
     This material goes through a Mott insulator to metal transition
     as the doping by P increases. At $n_c \sim 3.7 \ 10^{18} $ P per
     cm$^3$ there is a metallic state. The spin susceptibility of
     Si:P gets enhanced and becomes strongly temperature dependent as
     the metal-insulator transition is approached from the metallic
     side. The T-dependence observed does not fit the $ 1/T $
     behavior expected for weakly interacting localized spins either.
     Moreover, the spin-lattice relaxation times in barely metallic
     Si:P is strongly temperature dependent, $ T_1^{-1} \sim T^{-1/2}
     $, similarly the correction to the zero-T conductivity, $ \sigma
     \sim T ^{1/2} $, in this material. There are theories which
     associate these anomalies to spin fluctuations induced due to
     incipient localization. There is a subtle interplay of disorder
     and correlation effect in this material. Only spin fluctuation
     kind of theory will not work.  

     Some transition metal (and also some actinide) inter-metallic
     compounds show a low saturation moment per transition metal atom
     and a low magnetic transition temperature $\rm T_c$ compared
     with conventional ferromagnetic materials like Fe, Co and Ni.
     These compounds are known as weak itinerant electron
     ferromagnets. The prototype examples are, $\rm ZrZn_2$, 
     $\rm Ni_3Al$, and $\rm Sc_3In$. \cite{SGM90} Their low
     temperature properties have been discussed within spin
     fluctuation theories for a long time. \cite{Mo85}
     Here we compare the specific heat behavior of Sc$_3$In above $
     T_c $ with our present calculation  (Fig.\ (\ref{sc3in})). The 
     experimental curves are
     due to Ikeda \cite{Ike91} and show a good fit to the theory with
     $ \Delta \rho (T) \sim T $, and $ C_v(T) / T \sim ( T - T_c )
     ^{-0.25} $.  

     The example of MnSi is interesting from the perspective of the
     present work. The material has a transition temperature around
     $30 K $. As the hydrostatic pressure is applied the $T_c$
     decreases continuously and collapses towards absolute zero at
     $p_c = 14.6 kbar$. This is an example where an approach to a
     zero temperature quantum phase transition can be observed as a
     function of pressure. This has been done by the Cambridge group.
     \cite{Pfl94} The deviation of the resistivity curve from the
     $T^2$ behavior gets pronounced as $p_c$ is approached. In Fig.\
     (\ref{mnsi})we have compared $\Delta \rho /T^2 $ with a power
     law temperature dependence as suggested in our calculation.  

     We have compared the nuclear spin relaxation rate of $^{27}Al$
     in $ Ni_3Al $ as a function of temperature.\cite{Mas83} This
     material has a transition temperature about 41 K and shows all
     other characteristic properties of weak itinerant ferromagnet.
     \cite{Sasa83} The low field data fit to power law $T^{-0.89}$
     for $(T_1T)^{-1} $ as shown in Fig.\ (\ref{ni3al}).

     The heavy fermion material CeCu$_6$ is non-magnetic. On alloying
     with Au the lattice expands and an antiferromagnetic order is
     observed in CeCu$_{6-x}$Au$_x$ above a critical concentration
     $x_c \approx 0.1 $. The N\`eel temperature of the
     anti-ferromagnetic heavy-fermion alloy CeCu$_{5.7}$Au$_{0.3}$ 
     can be continuously tuned to zero with increasing hydrostatic
     pressure. At the critical pressure the specific heat has been
     fitted to $C/T \sim \ln T_0/T $ curve.\cite{Bog95} We analyze
     the data again and fit the curve to our prediction ($T^{0.58}$
     corresponding to the temperature range of interest) in the
     Fig.\ (\ref{cecuau}).  

     \section{conclusion}
     We have calculated the temperature dependence of various
     physical properties near the quantum phase transition point. The
     results hold for electronic phase transitions with a finite
     $T_c$ also. This is clear, as the results from the first rows
     in Table (1) match with some well known results in literature.
     However, they are pronounced and a clear non-Fermi liquid
     behavior is obtained when $T_c \rightarrow 0$. Our results are
     perturbative, but as discussed in the text the fluctuation
     correlation term is always smaller than the mean fluctuation
     field term.  The behavior of these quantities is different in
     ferromagnet from the antiferromagnetic system. This is a
     reflection of the fact that the order parameter fluctuations
     have different form of dispersion in these systems. Finally we
     made some remarks about the inclusion of the effect of disorder
     near quantum critical point within the spin fluctuation
     formalism. The present approach can be applied to other systems
     also. One only needs an appropriate form of the order parameter
     correlation function to calculate various quantities. For
     example this approach can be applied to systems with a pseudo
     gap \cite{KS90} in the excitation spectrum and also with phonon
     like dispersion as it happens in short coherence length
     superconductors. For example in 2D short coherence length
     superconductors it has been shown through Monte Carlo
     simulations that the relaxation rate varies as the spin
     susceptibility, \cite{Nan95} $(\alpha(T))^{-1}$ of the system,
     which matches with our result on 2D antiferromagnets.  
     
     \newpage
     
     \centerline {\bf FIGURE CAPTIONS}

\begin{figure}
\caption{Self energy diagrams for the spin fluctuation propagator.
\label{feyndiag1}}
\end{figure}

\begin{figure}
     \caption{Self energy in the mean fluctuation field
              approximation
\label{feyndiag2}} 
\end{figure}

\begin{figure}
     \caption{Schematic diagram of the various temperature scales
     involved in the disordered material
\label{line}}
\end{figure}

     \begin{figure}
     \caption{Plot of $C_v/T$ as a function of $T-T_c$ for $Sc_3In$.
     The experimental points are represented by circles and the solid
     line represents the theoretical fit.
     $\delta=1/12,\gamma=1/2,T_F\approx 1000K$. 
\label{sc3in}}
\end{figure} 

\begin{figure}
     \caption{Plot of $\rho(T)$ as a function of $T$ for MnSi.
     The experimental points are represented by circles and the solid
     line represents the theoretical fit. 
$\delta=1/12,\gamma=1/2,T_F\approx 1000K$.
\label{mnsi}}
\end{figure}

\begin{figure}
     \caption{Plot of $(T_1T)^{-1}$ as a function of $T$ for
     $Ni_3Al$. The experimental points are represented by circles and
     the solid line represents the theoretical fit. 
$\delta=1/12,\gamma=1/2,T_F\approx 1000K$.
\label{ni3al}}
\end{figure}

\begin{figure}
     \caption{Plot of $C_v$ as a function of $T$ for
     $CeCu_{5.7}Au_{0.3}$ at a pressure of 8.2 kbar. The experimental
     points are represented by circles and the solid line represents
     the theoretical fit. $\delta=1/2\pi,\gamma=1,T_F\approx 5K$.
\label{cecuau}}
\end{figure}

     \centerline {\bf TABLE CAPTIONS}

\begin{table}
\caption{ Summary of the temperature dependence of various thermal and 
     transport properties near a quantum phase transition point. The
     first row gives the non-selfconsistent calculation scheme. For
     ferromagnets the upper cutoff for $q$ is $T^{1/3}$ while for
     antiferromagnets it is $T^{1/2}$. The second row gives the 
     selfconsistent calculation scheme results. The upper cutoff for
     $q$ has been taken to be 1. The range of temperatures in which
     these exponents have been calculated is $T = 10^{-3}$ to
     $10^{-2}$. The third row gives the classical spin fluctuation
     results. (i.e. the first row with $q_T$ as a constant.) }
\end{table}

\end{document}